\documentclass[12pt]{article}\usepackage[hyperfootnotes=false]{hyperref}
\usepackage{epsfig}
\usepackage{float}
\usepackage{empheq}
\usepackage{bbold}

\usepackage{xspace}

\makeatletter

\@addtoreset{equation}{section}
\makeatother

\usepackage[utf8]{inputenc}
\usepackage{amsmath}
\usepackage{caption}
\usepackage{amsmath}
\usepackage{amssymb}
\usepackage{graphicx}

\newlength{\abstractwidth}
\renewcommand{\thefootnote}{\fnsymbol{footnote}}
\renewcommand{\thanks}[1]{\footnote{#1}} 
\newcommand{\starttext}{
\setcounter{footnote}{0}
\renewcommand{\thefootnote}{\arabic{footnote}}}

\newcommand{\be}{\begin{equation}}
\newcommand{\bea}{\begin{eqnarray}}
\newcommand{\eea}{\end{eqnarray}}
\newcommand{\beq}{\begin{equation}}
\newcommand{\ee}{\end{equation}}

	\newcommand*\widefbox[1]{\fbox{\hspace{2em}#1\hspace{2em}}}

	\def\dsp.{de Sitter space.}
	\def\eq{&=&}

	\def\simleq{\; \raise0.3ex\hbox{$<$\kern-0.75em
			\raise-1.1ex\hbox{$\sim$}}\; }
	\def\simgeq{\; \raise0.3ex\hbox{$>$\kern-0.75em
			\raise-1.1ex\hbox{$\sim$}}\; }
	
	\def\bi{\begin{itemize}}
		\def\ei{\end{itemize}}

	\def\bsub{ \begin{subequations}
			\begin{empheq}[box=\widefbox]{align}  }
			\def\esub{ \end{empheq}
	\end{subequations}}
	
	\def\1{\(  \mathbb{1} \)}

	\def\bn{\bigskip \noindent}

	\def\dk{${\rm DSSYK_{\infty}}$}
	
	\makeatletter
	\g@addto@macro\normalsize{%
		\setlength\abovedisplayskip{10pt}
		\setlength\belowdisplayskip{20pt}
		\setlength\abovedisplayshortskip{10pt}
		\setlength\belowdisplayshortskip{20pt}
	}
	\makeatother
	
	\usepackage{color}

	\usepackage{authblk}
	\title{\Large \bf Where is the entropy in DSSYK--de Sitter? Correction to a Wrong Claim }

	\author[1,2]{\Large Leonard Susskind}
	\affil[1]{LITP and Department of Physics, Stanford University, Stanford, CA 94305-4060, USA \vspace{1em}}
	\affil[2]{Google, Mountain View, CA, USA}
	\date{}
	\usepackage{upgreek}

\begin{document}
		
		\begin{titlepage}
			\maketitle
			
			\begin{abstract}

\end{abstract}

		\end{titlepage}
		
		\rightline{}
		\bigskip
		\bigskip\bigskip\bigskip\bigskip
		\bigskip
		
		\starttext \baselineskip=17.63pt \setcounter{footnote}{0}

	\LARGE
		
		\tableofcontents
		
	\section{Notational Issues and Scales}
		
Before discussing main point of this note I'll address some notational issues. I will use the notations of my Stoney Brook lectures with one exception\footnote{Here is the URL to access the lectures.
	
	\normalsize

 $$https://scgp.stonybrook.edu/video_portal
 /results.php?profile_id=2366$$
	}.
	 \LARGE
	 
	 \subsection*{Coupling Constants}
	In earlier papers \cite{Sekino:2025bsc}\cite{Miyashita:2025rpt}
	and in my lectures  I identified the \dk \ parameter $\lambda$ with $g_{string} $ by which I meant the closed-string coupling constant. 
	This may be confusing because the 't Hooft model  \cite{tHooft:1974pnl}
	 does not have closed strings; it is a pure open string theory. What I should have said is that 
$$
	\lambda = g_{string}^2
$$
	where $g_{string }$ is the open-string coupling. But it's too late for that now. Instead I will write,
	\be 
	\lambda = g_{open}^2 
	\label{defs}
	\ee
As I said, there are no closed strings in the 't Hooft model so that 
\be 
g_{closed} =g_{open}^2
\label{cl-op2}
\ee 
is merely a definition,  but
 in string theory with both open and closed strings \eqref{cl-op2}  correctly relates closed and open string couplings.

\bn

\subsection*{Review of Scales}
\begin{figure}[H]
		\begin{center}
			\includegraphics[scale=.5]{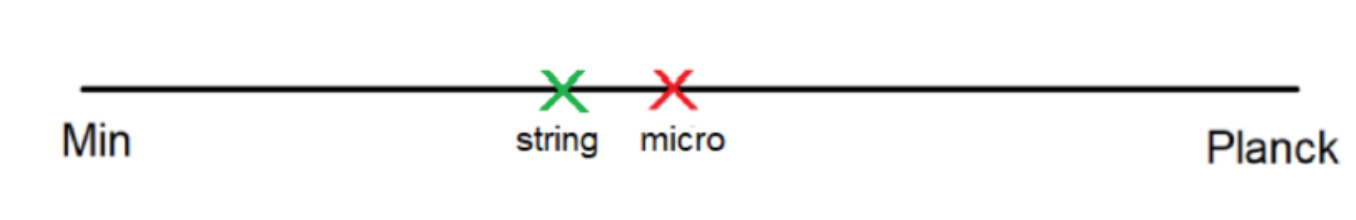}
			\caption{Energy scales shown from the lowest  $M_{min},$} i,e,. the Gibbons Hawking temperature, to the maximum mass---the scale at which the conical deficit is equal to $2\pi.$ The micro scale is the geometric mean of $M_{max}$ and $M_{min}.$ 
			\label{scales}
		\end{center}
	\end{figure}

In figure \ref{scales} the relation between the various scales that appeared in my lectures is shown on a logarithmic plot  \cite{Susskind:2022bia}. 
	The relations between the  scales are,
	\bea  
	M_{micro} &=& \sqrt{M_{planck}M_{min}} =  M_{Planck}/\sqrt{N} \cr \cr
	  \frac{M_{planck}}{M_{min} }  \eq N  \cr \cr
	M_{string} \eq \sqrt{\lambda} M_{micro} =  \sqrt{\frac{\lambda}{N}    } M_{Planck}
	\label{scls}
	\eea

\section{The Phase Boundary}\label{Corrections}
In this note I want to correct a serious error that I made in a number of papers \cite{Sekino:2025bsc}\cite{Miyashita:2025rpt} as well as lectures at the Simons Center in Stoney Brook. The error has to do with the nature of the stretched horizon in the \dk/ JT-de Sitter correspondence. 
 It relates to a puzzle that bothered me, having to do with the role of the  Planck scale in the \dk/ JT-de Sitter correspondence. In my lectures the string  scale played a  very prominent role but the Planck length $\ell_{planck} $ played almost none. This was based on a misconception that I will explain shortly.
 
\begin{figure}[H]
		\begin{center}
			\includegraphics[scale=.2]{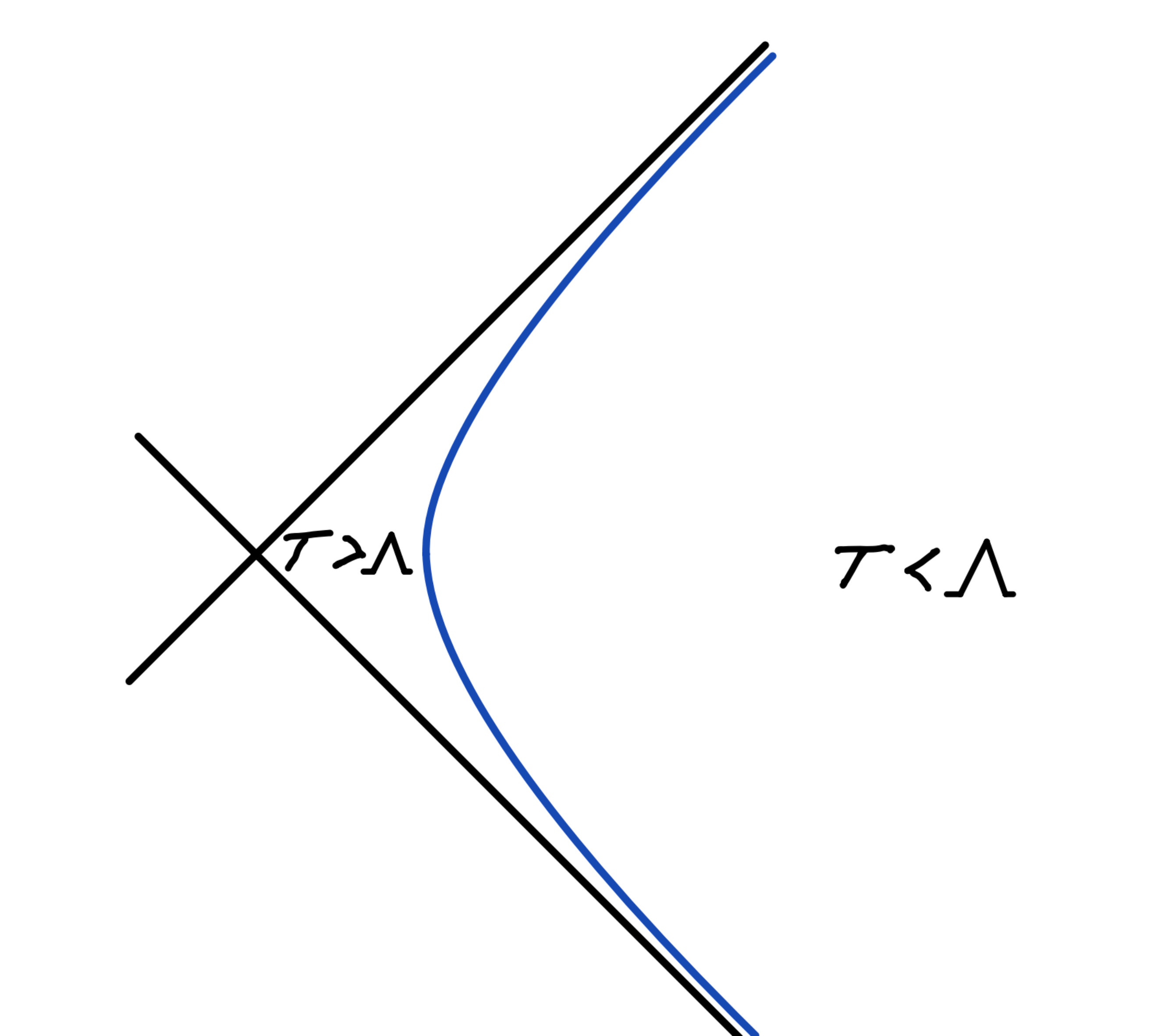}
			\caption{Rindler space divided into hot and cold regions by a curve along which the temperature is the QCD-scale. The distance of the dividing curve from the horizon is the string length. The mistake was to identify the curve with the phase boundary separating the cold confined region from the hot plasma deconfined region.}
			\label{rindler}
		\end{center}
	\end{figure}

 Let me summarize what I said at the end of Lecture 2 (and then repeated at the end of Lecture 4) where I was explaining the flat-space limit, in which the static patch becomes the Rindler patch. I began by dividing the Rindler patch into two regions, a cold region $T<\Lambda$ and a hot region $T>\Lambda $ (see figure \ref{rindler}). $\Lambda$ is the QCD scale\footnote{The QCD scale $\Lambda$ should not be confused with the DSSYK parameter $\lambda.$} defined by the string tension $\tau = \Lambda^2$. It also corresponds to the string-mass $M_{string }$ in equations \eqref{scls}.

The mistake was to conflate $\Lambda$ with the confinement-deconfinement transition temperature $T_c$ which would be correct in four-dimensional QCD but not in two-dimensional QCD. I also mistakenly identified the blue curve, at a distance $\ell_{string}$ from the horizon, as the phase boundary separating the confined and deconfined phases.

 My conclusion was that in the region
 $\rho< \ell_{string}$ 
 ($\rho $ is the proper distance from the horizon) quarks are unconfined and propagate freely,  and that the entropy stored in this stringy region is order $N,$ while the entropy in the cold region is order $1.$ 
 
 The argument is correct except for the fact that the confinement-deconfinement transition in $QCD_2$   is not at  the string-scale temperature $\Lambda,$ but at the parametrically higher temperature\footnote{I thank Steve Shenker and Sumit Das for pointing this out to me.} \cite{McLerran:1985uh},
 \bea 
 T_c &=& \Lambda \sqrt{N} \cr \cr
 &=&  M_{string}\sqrt{N}
 \label{Tc}
 \eea 
Using the relations \eqref{scls}
 \bea
 M_{string} &=& \sqrt{\lambda}M_{micro} \cr \cr
 \eq  \sqrt{\frac{\lambda}{N}}M_{planck}
 \label{Ms}
 \eea
 and $\lambda = g_{open}^2 $
 we find,
\be 
T_c = g_{open} M_{planck}
\label{Tc=gMp}
\ee
rather than the parametrically lower scale $M_{string}.$ Correspondingly, the distance of the stretched horizon from the mathematical horizon is,
\be  
\rho_{sh} = T_c^{-1}  = g_{open}^{-1} \ \ell_{planck}
\ee

Assuming $g_{open} \sim 1$ the Plank scale  emerges as    the scale at which the horizon entropy is stored. Perhaps this is not unexpected in retrospect, but it is interesting the way it follows from the the statistical mechanics of the 't Hooft model as calculated by McLerran and Sen.

\section{Comparison with String Theory}

Let us compare \eqref{Tc=gMp}  with closed string theory in higher dimensions where there 
 really is a transition at $T_c = M_{string},$  namely the Hagedorn transition. However, it is also true that the string and Planck scales are parametrically the same,
 \be  
 M_{string} = g_{closed}M_{planck}.
 \label{4da}
 \ee
 Thus it follows  in string theory that,
 \be
 T_c = g_{closed}M_{planck} 
 \label{4db}
 \ee
 Equations \eqref{Tc=gMp} and \eqref{4db}     are remarkably similar with the exception that $g_{open}$ is replaced by $g_{closed}$. That is not at all surprising since the 't Hooft model is a pure open string theory with no closed strings in its spectrum.
 
 \section{Summary and Conclusion}
It was in  error on my part when I  assumed that the confinement-deconfinement temperature was $\Lambda,$ or equivalently the string scale, in the 't Hooft model. In fact the transition temperature was shown by McLerran and Sen to be  much higher at
 $$T_c = \Lambda \sqrt{N}.$$ That translates to a temperature near the Planck scale, $$T_c = g_{open}M_{planck}.$$ This parallels closed  string theory where the corresponding formula is $$T_c = g_{closed}M_{planck}.$$
 
 One lesson is that the  huge horizon entropy in both cases is a Planck-scale phenomenon. In higher dimensional string theory there is not much difference between the string and Planck scales but in the two-dimensional
  case the difference is very large: the Planck and string scales differ by a factor of order $\sqrt{N}.$
 
 In both  cases the transition temperature has the form
 $$T_c = g M_{planck}$$ with $g$ being the open string coupling for the 't Hooft model, and the closed string coupling for  closed string theory. This reflects the fact that the 't Hooft model is a theory of open strings.

\section*{Acknowledgement}
I would would like to thank Steve Shenker and Sumit Das for pointing out that there is no phase transition in the 't Hooft model at the QCD scale. Sumit called my attention to the paper by McLerran and Sen.


\begin{thebibliography}{99} 

\bibitem{Sekino:2025bsc}
Y. ~Sekino and L. ~Susskind,
``Double-Scaled SYK, QCD, and the Flat Space Limit of de Sitter Space,''
[arXiv:2501.09423 [hep-th]].


\bibitem{Miyashita:2025rpt}
"Miyashita, Shoichiro and Sekino, Yasuhiro and Susskind, Leonard",
 "{DSSYK at Infinite Temperature: The Flat-Space Limit and the 't Hooft Model}",
    eprint = "2506.18054",
   
\bibitem{tHooft:1974pnl}
G.~'t~Hooft, 
``A Two-Dimensional Model for Mesons,''
Nucl. Phys. B \textbf{75}, 461-470  (1974)   
   
\bibitem{Susskind:2022bia}
L.~Susskind,
``De Sitter Space, Double-Scaled SYK, and the Separation of Scales in the Semiclassical Limit,''
JHAP \textbf{5}, 1, 1-30  (2025) 
[arXiv:2209.09999 [hep-th]].
    
   
			
		





   
	\bibitem{McLerran:1985uh}		
	 "McLerran, Larry D. and Sen, Ashoke",
 "{Thermodynamics of QCD in the Large N Limit}",
    reportNumber = "FERMILAB-PUB-85-055-T",
    doi = "10.1103/PhysRevD.32.2794",
   



    


		
	
			
		\end{thebibliography}
	\end{document}